\documentclass[amsmath,amssymb,aps,prd,11pt,tightenlines,superscriptaddress,nofootinbib,preprintnumbers,notitlepage]{revtex4-1}

\newcommand{\PRE}[1]{{#1}} 

\usepackage{subfigure}
\usepackage{orcidlink}

\usepackage{amsmath,amssymb,amsthm,amsfonts}
\usepackage{graphicx,tabularx}
\usepackage{color}
\usepackage{multirow}
\usepackage{comment}
\usepackage{enumitem}
\usepackage{cleveref}
\usepackage{slashed}
\usepackage[normalem]{ulem}
\usepackage{scalerel}
\usepackage{wrapfig}
\usepackage{cancel}
\usepackage{xcolor}

\newcommand{\be}{\begin{equation} \begin{aligned}}
\newcommand{\ee}{\end{aligned} \end{equation}}
\newcommand{\beqa}{\begin{eqnarray}}
\newcommand{\eeqa}{\end{eqnarray}}

\def\figureautorefname~#1\null{Fig.\,#1\null}
\def\tableautorefname~#1\null{Tab.\,#1\null}
\def\equationautorefname~#1\null{Eq.\,(#1)\null}

\newcommand{\mev}{\text{MeV}}
\newcommand{\gev}{\text{GeV}}
\newcommand{\tev}{\text{TeV}}

\newcommand{\fb}{\text{fb}}
\newcommand{\ifb}{\text{fb}^{-1}}

\newcommand{\m}{\text{m}}

\RequirePackage[normalem]{ulem}

\crefname{section}{Sec.}{Secs.}
\crefname{figure}{Fig.}{Figs.}
\crefname{equation}{Eq.}{Eqs.}
\crefname{table}{Table}{Tables}
\crefname{appendix}{Appendix}{Appendices}

\begin{document}

\preprint{CERN-EP-2023-161}

\title{\Large{Search for Dark Photons with the FASER detector at the LHC}
\PRE{\vspace*{0.1in} \\    
FASER Collaboration}
}

\author{\vspace{-0.12in} Henso Abreu\,\orcidlink{0000-0002-1599-2896}} 
\affiliation{Department of Physics and Astronomy, Technion---Israel Institute of Technology, Haifa 32000, Israel}

\author{John Anders\,\orcidlink{0000-0002-1846-0262}} 
\affiliation{CERN, CH-1211 Geneva 23, Switzerland}

\author{Claire Antel\,\orcidlink{0000-0001-9683-0890}} 
\affiliation{D\'epartement de Physique Nucl\'eaire et Corpusculaire, University of Geneva, CH-1211 Geneva 4, Switzerland}

\author{Akitaka Ariga\,\orcidlink{0000-0002-6832-2466}} 
\affiliation{Albert Einstein Center for Fundamental Physics, Laboratory for High Energy Physics, University of Bern, Sidlerstrasse 5, CH-3012 Bern, Switzerland}
\affiliation{Department of Physics, Chiba University, 1-33 Yayoi-cho Inage-ku, 263-8522 Chiba, Japan}

\author{Tomoko Ariga\,\orcidlink{0000-0001-9880-3562}} 
\affiliation{Kyushu University, Nishi-ku, 819-0395 Fukuoka, Japan}

\author{Jeremy Atkinson\,\orcidlink{0009-0003-3287-2196}}    
\affiliation{Albert Einstein Center for Fundamental Physics, Laboratory for High Energy Physics, University of Bern, Sidlerstrasse 5, CH-3012 Bern, Switzerland}

\author{Florian~U.~Bernlochner\,\orcidlink{0000-0001-8153-2719}} 
\affiliation{Universit\"at Bonn, Regina-Pacis-Weg 3, D-53113 Bonn, Germany}

\author{Tobias Boeckh\,\orcidlink{0009-0000-7721-2114}} 
\affiliation{Universit\"at Bonn, Regina-Pacis-Weg 3, D-53113 Bonn, Germany}

\author{Jamie Boyd\,\orcidlink{0000-0001-7360-0726}} 
\affiliation{CERN, CH-1211 Geneva 23, Switzerland}

\author{Lydia Brenner\,\orcidlink{0000-0001-5350-7081}} 
\affiliation{Nikhef National Institute for Subatomic Physics, Science Park 105, 1098 XG Amsterdam, Netherlands}

\author{Franck Cadoux} 
\affiliation{D\'epartement de Physique Nucl\'eaire et Corpusculaire, University of Geneva, CH-1211 Geneva 4, Switzerland}

\author{David~W.~Casper\,\orcidlink{0000-0002-7618-1683}} 
\affiliation{Department of Physics and Astronomy, University of California, Irvine, CA 92697-4575, USA}

\author{Charlotte Cavanagh\,\orcidlink{0009-0001-1146-5247}} 
\affiliation{University of Liverpool, Liverpool L69 3BX, United Kingdom}

\author{Xin Chen\,\orcidlink{0000-0003-4027-3305}} 
\affiliation{Department of Physics, Tsinghua University, Beijing, China}

\author{Andrea Coccaro\,\orcidlink{0000-0003-2368-4559}} 
\affiliation{INFN Sezione di Genova, Via Dodecaneso, 33--16146, Genova, Italy}

\author{Monica D’Onofrio\,\orcidlink{0000-0003-2408-5099}} 
\affiliation{University of Liverpool, Liverpool L69 3BX, United Kingdom}

\author{Ansh Desai\,\orcidlink{0000-0002-5447-8304}} 
\affiliation{University of Oregon, Eugene, OR 97403, USA}

\author{Sergey Dmitrievsky\,\orcidlink{0000-0003-4247-8697}} 
\affiliation{Affiliated with an international laboratory covered by a cooperation agreement with CERN.}

\author{Candan Dozen\,\orcidlink{0000-0002-4301-634X}}
\affiliation{Institut de Physique des 2 Infinis de Lyon (IP2I), Villeurbanne, France}

\author{Yannick Favre}
\affiliation{D\'epartement de Physique Nucl\'eaire et Corpusculaire, University of Geneva, CH-1211 Geneva 4, Switzerland}

\author{Deion Fellers\,\orcidlink{0000-0002-0731-9562}} 
\affiliation{University of Oregon, Eugene, OR 97403, USA}

\author{Jonathan~L.~Feng\,\orcidlink{0000-0002-7713-2138}} 
\affiliation{Department of Physics and Astronomy, University of California, Irvine, CA 92697-4575, USA}

\author{Carlo Alberto Fenoglio\,\orcidlink{0009-0007-7567-8763}}    
\affiliation{D\'epartement de Physique Nucl\'eaire et Corpusculaire, University of Geneva, CH-1211 Geneva 4, Switzerland}

\author{Didier Ferrere\,\orcidlink{0000-0002-5687-9240}} 
\affiliation{D\'epartement de Physique Nucl\'eaire et Corpusculaire, University of Geneva, CH-1211 Geneva 4, Switzerland}

\author{Iftah Galon\,\orcidlink{0000-0003-0104-231X}}
\noaffiliation

\author{Stephen Gibson\,\orcidlink{0000-0002-1236-9249}} 
\affiliation{Royal Holloway, University of London, Egham, TW20 0EX, United Kingdom}

\author{Sergio Gonzalez-Sevilla\,\orcidlink{0000-0003-4458-9403}} 
\affiliation{D\'epartement de Physique Nucl\'eaire et Corpusculaire, University of Geneva, CH-1211 Geneva 4, Switzerland}

\author{Yuri Gornushkin\,\orcidlink{0000-0003-3524-4032}} 
\affiliation{Affiliated with an international laboratory covered by a cooperation agreement with CERN.}

\author{Carl Gwilliam\,\orcidlink{0000-0002-9401-5304}} 
\affiliation{University of Liverpool, Liverpool L69 3BX, United Kingdom}

\author{Daiki Hayakawa\,\orcidlink{0000-0003-4253-4484}} 
\affiliation{Department of Physics, Chiba University, 1-33 Yayoi-cho Inage-ku, 263-8522 Chiba, Japan}

\author{Shih-Chieh Hsu\,\orcidlink{0000-0001-6214-8500}} 
\affiliation{Department of Physics, University of Washington, PO Box 351560, Seattle, WA 98195-1460, USA}

\author{Zhen Hu\,\orcidlink{0000-0001-8209-4343}} 
\affiliation{Department of Physics, Tsinghua University, Beijing, China}

\author{Giuseppe Iacobucci\,\orcidlink{0000-0001-9965-5442}} 
\affiliation{D\'epartement de Physique Nucl\'eaire et Corpusculaire, University of Geneva, CH-1211 Geneva 4, Switzerland}

\author{Tomohiro Inada\,\orcidlink{0000-0002-6923-9314}} 
\affiliation{Department of Physics, Tsinghua University, Beijing, China}

\author{Sune Jakobsen\,\orcidlink{0000-0002-6564-040X}} 
\affiliation{CERN, CH-1211 Geneva 23, Switzerland}

\author{Hans Joos\,\orcidlink{0000-0003-4313-4255}}    
\affiliation{CERN, CH-1211 Geneva 23, Switzerland}
\affiliation{II.~Physikalisches Institut, Universität Göttingen, Göttingen, Germany}

\author{Enrique Kajomovitz\,\orcidlink{0000-0002-8464-1790}} 
\affiliation{Department of Physics and Astronomy, Technion---Israel Institute of Technology, Haifa 32000, Israel}

\author{Hiroaki Kawahara\,\orcidlink{0009-0007-5657-9954}} 
\affiliation{Kyushu University, Nishi-ku, 819-0395 Fukuoka, Japan}

\author{Alex Keyken}    
\affiliation{Royal Holloway, University of London, Egham, TW20 0EX, United Kingdom}

\author{Felix Kling\,\orcidlink{0000-0002-3100-6144}} 
\affiliation{Deutsches Elektronen-Synchrotron DESY, Notkestr. 85, 22607 Hamburg, Germany}

\author{Daniela Köck\,\orcidlink{0000-0002-9090-5502}} 
\affiliation{University of Oregon, Eugene, OR 97403, USA}

\author{Umut Kose\,\orcidlink{0000-0001-5380-9354}} 
\affiliation{CERN, CH-1211 Geneva 23, Switzerland}

\author{Rafaella Kotitsa\,\orcidlink{0000-0002-7886-2685}} 
\affiliation{CERN, CH-1211 Geneva 23, Switzerland}

\author{Susanne Kuehn\,\orcidlink{0000-0001-5270-0920}} 
\affiliation{CERN, CH-1211 Geneva 23, Switzerland}

\author{Helena Lefebvre\,\orcidlink{0000-0002-7394-2408}} 
\affiliation{Royal Holloway, University of London, Egham, TW20 0EX, United Kingdom}

\author{Lorne Levinson\,\orcidlink{0000-0003-4679-0485}} 
\affiliation{Department of Particle Physics and Astrophysics, Weizmann Institute of Science, Rehovot 76100, Israel}

\author{Ke Li\,\orcidlink{0000-0002-2545-0329}} 
\affiliation{Department of Physics, University of Washington, PO Box 351560, Seattle, WA 98195-1460, USA}

\author{Jinfeng Liu}
\affiliation{Department of Physics, Tsinghua University, Beijing, China}

\author{Jack MacDonald\,\orcidlink{0000-0002-3150-3124}}    
\affiliation{Institut f\"ur Physik, Universität Mainz, Mainz, Germany}

\author{Chiara Magliocca\,\orcidlink{0009-0009-4927-9253}} 
\affiliation{D\'epartement de Physique Nucl\'eaire et Corpusculaire, University of Geneva, CH-1211 Geneva 4, Switzerland}

\author{Fulvio Martinelli\,\orcidlink{0000-0003-4221-5862}} 
\affiliation{D\'epartement de Physique Nucl\'eaire et Corpusculaire, University of Geneva, CH-1211 Geneva 4, Switzerland}

\author{Josh McFayden\,\orcidlink{0000-0001-9273-2564}} 
\affiliation{Department of Physics \& Astronomy, University of Sussex, Sussex House, Falmer, Brighton, BN1 9RH, United Kingdom}

\author{Sam Meehan\,\orcidlink{0000-0002-3613-7514}}
\affiliation{CERN, CH-1211 Geneva 23, Switzerland}
\affiliation{Science and Technology Policy Fellow at the American Association for the Advancement of Science}

\author{Matteo Milanesio\,\orcidlink{0000-0001-8778-9638}} 
\affiliation{D\'epartement de Physique Nucl\'eaire et Corpusculaire, University of Geneva, CH-1211 Geneva 4, Switzerland}

\author{Théo Moretti\,\orcidlink{0000-0001-7065-1923}} 
\affiliation{D\'epartement de Physique Nucl\'eaire et Corpusculaire, University of Geneva, CH-1211 Geneva 4, Switzerland}

\author{Magdalena Munker\,\orcidlink{0000-0003-2775-3291}} 
\affiliation{D\'epartement de Physique Nucl\'eaire et Corpusculaire, University of Geneva, CH-1211 Geneva 4, Switzerland}

\author{Mitsuhiro Nakamura}
\affiliation{Nagoya University, Furo-cho, Chikusa-ku, Nagoya 464-8602, Japan}

\author{Toshiyuki Nakano}
\affiliation{Nagoya University, Furo-cho, Chikusa-ku, Nagoya 464-8602, Japan}

\author{Friedemann Neuhaus\,\orcidlink{0000-0002-3819-2453}} 
\affiliation{Institut f\"ur Physik, Universität Mainz, Mainz, Germany}

\author{Laurie Nevay\,\orcidlink{0000-0001-7225-9327}} 
\affiliation{CERN, CH-1211 Geneva 23, Switzerland}
\affiliation{Royal Holloway, University of London, Egham, TW20 0EX, United Kingdom}

\author{Ken Ohashi}
\affiliation{Albert Einstein Center for Fundamental Physics, Laboratory for High Energy Physics, University of Bern, Sidlerstrasse 5, CH-3012 Bern, Switzerland}

\author{Hidetoshi Otono\,\orcidlink{0000-0003-0760-5988}} 
\affiliation{Kyushu University, Nishi-ku, 819-0395 Fukuoka, Japan}

\author{Hao Pang\,\orcidlink{0000-0002-1946-1769}} 
\affiliation{Department of Physics, Tsinghua University, Beijing, China}

\author{Lorenzo Paolozzi\,\orcidlink{0000-0002-9281-1972}} 
\affiliation{D\'epartement de Physique Nucl\'eaire et Corpusculaire, University of Geneva, CH-1211 Geneva 4, Switzerland}
\affiliation{CERN, CH-1211 Geneva 23, Switzerland}

\author{Brian Petersen\,\orcidlink{0000-0002-7380-6123}} 
\affiliation{CERN, CH-1211 Geneva 23, Switzerland}

\author{Markus Prim\,\orcidlink{0000-0002-1407-7450}} 
\affiliation{Universit\"at Bonn, Regina-Pacis-Weg 3, D-53113 Bonn, Germany}

\author{Michaela Queitsch-Maitland\,\orcidlink{0000-0003-4643-515X}} 
\affiliation{University of Manchester, School of Physics and Astronomy, Schuster Building, Oxford Rd, Manchester M13 9PL, United Kingdom}

\author{Hiroki Rokujo}
\affiliation{Nagoya University, Furo-cho, Chikusa-ku, Nagoya 464-8602, Japan}

\author{Elisa Ruiz-Choliz\,\orcidlink{0000-0002-2417-7121}} 
\affiliation{Institut f\"ur Physik, Universität Mainz, Mainz, Germany}

\author{Jorge Sabater-Iglesias\,\orcidlink{0000-0003-2328-1952}} 
\affiliation{D\'epartement de Physique Nucl\'eaire et Corpusculaire, University of Geneva, CH-1211 Geneva 4, Switzerland}

\author{Jakob Salfeld-Nebgen}
\affiliation{CERN, CH-1211 Geneva 23, Switzerland}

\author{Osamu Sato\,\orcidlink{0000-0002-6307-7019}} 
\affiliation{Nagoya University, Furo-cho, Chikusa-ku, Nagoya 464-8602, Japan}

\author{Paola Scampoli\,\orcidlink{0000-0001-7500-2535}} 
\affiliation{Albert Einstein Center for Fundamental Physics, Laboratory for High Energy Physics, University of Bern, Sidlerstrasse 5, CH-3012 Bern, Switzerland}
\affiliation{Dipartimento di Fisica ``Ettore Pancini'', Universit\`a di Napoli Federico II, Complesso Universitario di Monte S. Angelo, I-80126 Napoli, Italy}

\author{Kristof Schmieden\,\orcidlink{0000-0003-1978-4928}} 
\affiliation{Institut f\"ur Physik, Universität Mainz, Mainz, Germany}

\author{Matthias Schott\,\orcidlink{0000-0002-4235-7265}} 
\affiliation{Institut f\"ur Physik, Universität Mainz, Mainz, Germany}

\author{Anna Sfyrla\,\orcidlink{0000-0002-3003-9905}} 
\affiliation{D\'epartement de Physique Nucl\'eaire et Corpusculaire, University of Geneva, CH-1211 Geneva 4, Switzerland}

\author{Savannah Shively\,\orcidlink{0000-0002-4691-3767}} 
\affiliation{Department of Physics and Astronomy, University of California, Irvine, CA 92697-4575, USA}

\author{Yosuke Takubo\,\orcidlink{0000-0002-3143-8510}} 
\affiliation{Institute of Particle and Nuclear Studies, KEK, Oho 1-1, Tsukuba, Ibaraki 305-0801, Japan}

\author{Noshin Tarannum\,\orcidlink{0000-0002-3246-2686}} 
\affiliation{D\'epartement de Physique Nucl\'eaire et Corpusculaire, University of Geneva, CH-1211 Geneva 4, Switzerland}

\author{Ondrej Theiner\,\orcidlink{0000-0002-6558-7311}} 
\affiliation{D\'epartement de Physique Nucl\'eaire et Corpusculaire, University of Geneva, CH-1211 Geneva 4, Switzerland}

\author{Eric Torrence\,\orcidlink{0000-0003-2911-8910}} 
\affiliation{University of Oregon, Eugene, OR 97403, USA}

\author{Sebastian Trojanowski\,\orcidlink{0000-0003-2677-0364}}
\affiliation{Astrocent, Nicolaus Copernicus Astronomical Center Polish Academy of Sciences, ul.~Rektorska 4, 00-614, Warsaw, Poland}
\affiliation{National Centre for Nuclear Research, Pasteura 7, Warsaw, 02-093, Poland}

\author{Svetlana Vasina\,\orcidlink{0000-0003-2775-5721}} 
\affiliation{Affiliated with an international laboratory covered by a cooperation agreement with CERN.}

\author{Benedikt Vormwald\,\orcidlink{0000-0003-2607-7287}} 
\affiliation{CERN, CH-1211 Geneva 23, Switzerland}

\author{Di Wang\,\orcidlink{0000-0002-0050-612X}} 
\affiliation{Department of Physics, Tsinghua University, Beijing, China}

\author{Eli Welch\,\orcidlink{0000-0001-6336-2912}} 
\affiliation{Department of Physics and Astronomy, University of California, Irvine, CA 92697-4575, USA}

\author{Samuel Zahorec\,\orcidlink{0009-0000-9729-0611}}
\affiliation{CERN, CH-1211 Geneva 23, Switzerland}
\affiliation{Charles University, Faculty of Mathematics and Physics, Prague; Czech Republic}

\author{Stefano Zambito\,\orcidlink{0000-0002-4499-2545}} 
\affiliation{D\'epartement de Physique Nucl\'eaire et Corpusculaire, University of Geneva, CH-1211 Geneva 4, Switzerland}

\begin{abstract}
The FASER experiment at the LHC is designed to search for light, weakly-interacting particles produced in proton-proton collisions at the ATLAS interaction point that travel in the far-forward direction. The first results from a search for dark photons decaying to an electron-positron pair, using a dataset corresponding to an integrated luminosity of $27.0$\,$\mathrm{fb}^{-1}$ collected at center-of-mass energy $\sqrt{s} = 13.6$\,TeV in 2022 in LHC Run 3, are presented. No events are seen in an almost background-free analysis, yielding world-leading constraints on dark photons with couplings $\epsilon \sim 2 \times 10^{-5} - 1 \times 10^{-4}$ and masses $\sim 17~\mev - 70~\mev$. The analysis is also used to probe the parameter space of a massive gauge boson from a U(1)$_{B-L}$ model, with couplings $g_{B-L} \sim 5\times10^{-6} - 2\times10^{-5}$ and masses $\sim 15~\mev - 40~\mev$ excluded for the first time.
\end{abstract}

\maketitle

\begin{center}
\copyright~2023 CERN for the benefit of the FASER Collaboration.
Reproduction of this article or parts of it is allowed as specified in the CC-BY-4.0 license.     
\end{center}


\section{Introduction} 
\label{sec:introduction}

The existence of dark matter is strong evidence for new particles beyond the Standard Model (SM) of particle physics.  Dark matter may be composed of a single particle or of more than one kind of particle, and the dark matter particles may interact only through gravity or also through additional forces. Dark matter therefore motivates a rich variety of ideas for beyond-the-SM (BSM) physics, and new insights into the particle nature of dark matter are of great interest in both particle physics and astrophysics~\cite{Bertone:2004pz,Feng:2010gw}. 

Although dark matter is only known to interact through gravity, the identification of its particle properties will be possible only if it is detected via other interactions. Among the best-motivated possibilities are interactions with SM particles through renormalisable couplings.  If dark matter is a component of a dark sector that contains a U(1) electromagnetic force, the dark sector may interact through a renormalisable interaction of the form $F^{\mu\nu} F^D_{\mu\nu}$, where $F_{\mu\nu}$ and $F^D_{\mu\nu}$ are the electromagnetic field strength tensors of the SM and the dark sector, respectively.  As a result of this interaction, the dark gauge boson mixes with the SM gauge boson, leading to a new particle, the dark photon $A'$~\cite{Holdom:1985ag}. If dark photons are light and weakly (or feebly) interacting, they are long-lived particles (LLPs) and can be produced in large numbers in the proton-proton collisions at the Large Hadron Collider (LHC)~\cite{Evans:2008zzb}; in viable regions of dark photon parameter space that will be probed in this analysis, as many as $10^8$ dark photons could be produced~\cite{Feng:2017uoz}. They can then travel hundreds of meters and decay to pairs of charged particles, producing a spectacular signal of new physics.

Other considerations also motivate BSM physics with signals similar to the dark photon scenario.  For example, the accidental conservation of baryon number $B$ and (total) lepton number $L$ in the SM suggests that these conserved quantities may be linked not just to global, but to local gauge symmetries. A particularly well-motivated example is the gauge symmetry U(1)$_{B-L}$~\cite{Davidson:1978pm, Marshak:1979fm}, which is not only conserved classically, but is also free of quantum anomalies, once three sterile neutrinos are introduced to give neutrinos mass. This model predicts a new particle, the $B-L$ gauge boson $A'_{B-L}$. For masses in the MeV to GeV range and small $B-L$ gauge couplings ($\sim 10^{-5}$) up to $10^8$ $A'_{B-L}$ gauge bosons may be produced and travel hundreds of meters before decaying to pairs of SM particles with $B-L$ charge~\cite{Bauer:2018onh,FASER:2018eoc}.

FASER is a new LHC experiment designed to search for light, weakly-interacting particles, including dark photons, $B-L$ gauge bosons, and other long-lived particles~\cite{Feng:2017uoz, FASER:2018eoc, FASER:2022hcn}.  The FASER detector is located approximately 480 m from the ATLAS interaction point (IP1) along the beam collision axis line-of-sight (LOS). Because they interact very weakly, dark photons and other LLPs produced at IP1 can travel along the LOS, pass through $\sim$100 m of rock and concrete without interacting, and then decay in FASER.  At the same time, SM particles, except for muons and neutrinos, produced at the ATLAS IP will either be bent away by the LHC magnets or stopped in the rock and concrete.  FASER is therefore well suited to search for dark photons and many other light and weakly-interacting particles in a very low background environment. 

This study presents the results of a search for LLPs using the FASER detector and a dataset corresponding to an integrated luminosity of $27.0~\ifb$ collected at center-of-mass energy $\sqrt{s} = 13.6~\tev$ from September to November 2022 during Run 3 of the LHC. In particular, the scenario where LLPs are produced in LHC collisions, travel to the FASER detector, and then decay to electron-positron pairs, $pp \rightarrow \mathrm{LLP} \rightarrow e^+e^-$, is considered.

\section{Long-Lived Particles at FASER}
\label{sec:DPatFASER}

In this section, the parameter spaces of the dark photon and $B-L$ gauge boson models are defined and the dominant production and decay processes that determine the signal at FASER are described.

The properties of the dark photon are defined through the Lagrangian terms
\be
\mathcal{L} \supset \frac{1}{2} \, m_{A'}^2  A'^2
- \epsilon \, e \sum_f q_f A'^{\, \mu} \, \bar{f} \gamma_{\mu} f \ ,
\ee
where $m_{A'}$ is the dark photon's mass, $\epsilon$ is the dark photon's kinematic mixing parameter, and the sum is over all SM fermions $f$ with SM electric charge $q_f$.  The dark photon may also couple to additional particles in the dark sector, such as the dark matter particle $\chi$.

In this analysis, it is assumed that $m_{A'} < 2 m_{\chi}$ and that the dark photon decays visibly to SM particles. Thermal freeze-out is then determined by the processes $\chi \chi \leftrightarrow A' \leftrightarrow f \bar{f}$.  For light masses $m_{A'} \sim \mev - \gev$ and loop-induced or otherwise suppressed couplings $\epsilon \sim 10^{-6} - 10^{-3}$, the dark matter particle's thermal relic density is in the right range to be a significant fraction of cosmological dark matter~\cite{Boehm:2003hm,Pospelov:2007mp,Feng:2008ya}.  These values of $m_{A'}$ and $\epsilon$ are therefore cosmologically favoured and provide a well-defined thermal relic target in the dark photon parameter space for experimental searches.

At the LHC, with these thermal relic target parameters and in the parameter space where FASER has discovery potential, the dominant source of dark photons is SM meson decay and dark bremsstrahlung:
\begin{itemize}
\setlength\itemsep{-0.05in}
\item Neutral pion decay $\pi^0 \to A' \gamma$: This mode is accessible for $m_{A'} < m_{\pi^0} \simeq 135~\mev$. The branching fraction is $B(\pi^0 \to A' \gamma) = 2\epsilon^2 (1 -  m^2_{A'} /m_{\pi^0}^2 )^3 B(\pi^0 \to \gamma \gamma) $ where $B(\pi^0 \to \gamma \gamma) \simeq 0.99$~\cite{ParticleDataGroup:2020ssz}. 
\item Eta meson decay $\eta \to A' \gamma$: This mode is open for $m_{A'} < m_{\eta} \simeq 548~\mev$. The branching fraction is $B(\eta \to A' \gamma) = 2\epsilon^2 (1 - m^2_{A'} / m_\eta^2 )^3 B(\eta\to \gamma \gamma)$ where $B(\eta \to \gamma \gamma) \simeq 0.39$~\cite{ParticleDataGroup:2020ssz}. 
\item Dark bremsstrahlung $p p \to p p  A'$: In this process, a dark photon is emitted via initial or final state radiation from colliding protons in a coherent way. This mode is open for dark photon masses up to $\mathcal{O}(2~\gev)$~\cite{Feng:2017uoz}. 
\end{itemize}
These processes produce a high-intensity beam of dark photons in the far-forward direction along the beamline.  Neutral pion decay is typically the leading signal contribution, but $\eta$ decay can be comparable for $m_{A'} \sim 100~\mev$, and dark bremsstrahlung can be comparable near the boundary of FASER's sensitivity~\cite{Feng:2017uoz}.
Other production mechanisms include the decays of heavier mesons (such as $\eta'$ or $\omega$) and direct Drell-Yan production $q \bar{q} \to A'$, but these are subdominant and are neglected.

Once produced, dark photons then may travel a macroscopic distance, leading to a striking signal of high-energy particles far from the $pp$ interaction point. FASER's dark photon sensitivity is largely determined by its location. For $E_{A'} \gg m_{A'} \gg m_e$, the decay length for a dark photon with lifetime $\tau$ travelling at speed $\beta = v/c$ is~\cite{Feng:2017uoz}
\begin{equation}
\label{eq:ap_decay_length}
L = c \beta \tau \gamma  \approx (80~\m ) \left[ \frac{10^{-5}}{\epsilon} \right]^2 
\left[ \frac{E_{A'}}{\tev} \right] \left[ \frac{100~\mev}{m_{A'}} \right]^2 \ .
\end{equation}
For dark photons with TeV energies, FASER can be expected to be sensitive to parameter space with $\epsilon \sim 10^{-5}$ and $m_{A'} \sim 100$~MeV. For dark photon masses in the range $2 m_e < m_{A'} < 2 m_{\mu} \simeq 211~\mev$, dark photons decay to electrons with $B(A' \to e^+ e^-) \approx 100\%$.

In the $B-L$ model, the properties of the $B-L$ gauge boson $A'_{B-L}$ are determined by the Lagrangian terms~\cite{FASER:2018eoc}
\begin{equation}
\mathcal{L} \supset \frac{1}{2} \, m_{A'_{B-L}}^2  A'^{\,2}_{B-L}
- g_{B-L} \sum_f Q_{B-L}^f A'^{\, \mu}_{B-L} \, \bar{f}  \gamma_{\mu}  f \ ,
\end{equation}
where $Q_{B-L}^f$ is the $B-L$ charge of fermion $f$. The parameter space of this model is defined by the $B-L$ gauge boson's mass $m_{A'_{B-L}}$ and the $B-L$ gauge coupling $g_{B-L}$.  

The $A'_{B-L}$ gauge boson is produced in a similar manner to the dark photon, with light meson decays and dark bremsstrahlung the dominant production mechanisms; the production rates are proportional to $g_{B-L}^2$, compared to $\epsilon^2$ as in the dark photon model. The boson can decay to all kinematically accessible states that possess $B-L$ charge. In this analysis, the region of phase space which FASER is sensitive to is confined to the mass range $2 m_e < m_{A'_{B-L}} < 2 m_{\mu} \simeq 211~\mev$, where the possible decays are to electrons, SM neutrinos, and possibly sterile neutrinos.  It is assumed that sterile neutrinos have masses greater than half the $A'_{B-L}$ gauge boson mass, and so decays to sterile neutrinos are kinematically inaccessible. The visible signal from decays to electrons therefore has a branching fraction of $B(A'_{B-L} \to e^+e^-)\approx 40\%$.  If decays to sterile neutrinos are allowed, the visible branching fraction could be as low as $B(A'_{B-L} \to e^+e^-) \approx 25\%$, slightly reducing the search sensitivity, but not to a significant extent.

\section{The FASER Detector}
\label{sec:FASER}

The FASER detector, located approximately 480~m away from IP1 in the TI12 tunnel that connects the LHC with the Super Proton Synchotron (SPS), is aligned with the IP1 LOS. However, due to the crossing angle in IP1, the LOS is offset vertically by 6.5~cm with respect to the centre of the detector, which is properly accounted for in the simulation. The detector is described in detail in Ref.~\cite{FASER:2022hcn}; a brief description is given here. The FASER$\nu$ tungsten/emulsion detector is dedicated to neutrino measurements, and it is not used in this analysis, but the eight interaction lengths of tungsten suppress potential backgrounds. 
\cref{fig:detector_signal} presents a sketch of the detector. In this analysis, the detector components of interest are the 1.5~m long detector decay volume and the tracking spectrometer, both of which are immersed in a 0.57~T dipole magnetic field, as well as the scintillator system and the electromagnetic calorimeter. The active transverse area of the detector is defined by the circular magnet aperture with a radius of 10~cm. 

\begin{figure}[tbp]
\includegraphics[width=0.95\textwidth]{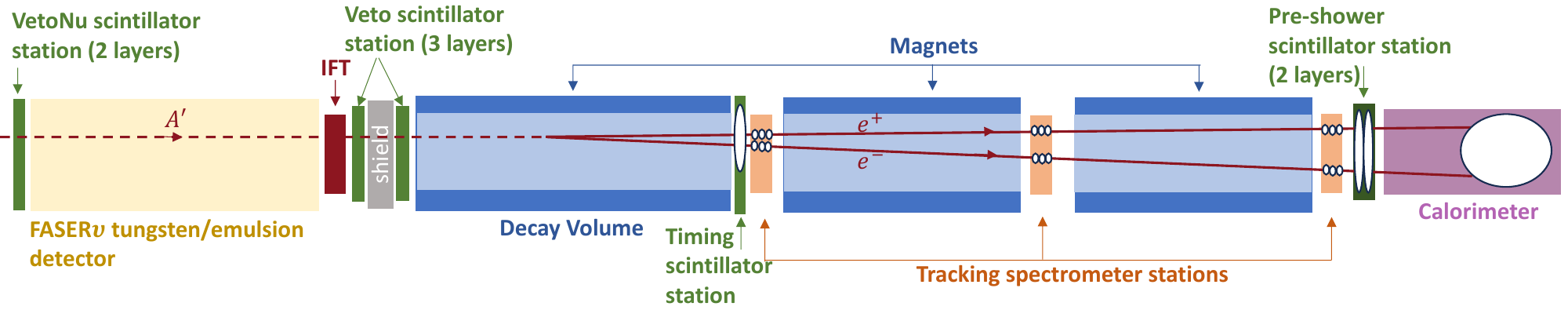}
\caption{A sketch presenting a side view of the FASER detector, showing the different detector systems as well as the signature of a dark photon ($A'$) decaying to an electron-positron pair inside the decay volume. The white blobs depict where measurements are taken for the $A'$ signal and the solid red lines represent the reconstructed tracks produced by the $e^+e^-$ pair.}   
\label{fig:detector_signal}
\end{figure}

The scintillator system is composed of four stations, each consisting of multiple scintillator counters. 
At the front of the detector is the VetoNu station, composed of two scintillator counters. Further downstream is the Veto station, constructed from three scintillator counters in front of the decay volume. Both the VetoNu and Veto stations have scintillators with a transverse size (30 $\times$ 35~cm$^2$ and 30 $\times$ 30~cm$^2$ respectively) significantly larger than the active region of the detector, which allows for the rejection of muons entering the detector at an angle with respect to the LOS.
The next scintillator station is the Timing station with two scintillator counters that separately cover the top and bottom half of the detector (with a small overlap) installed in front of the tracking spectrometer, used for triggering and timing measurements.
Finally, the Pre-shower station is in front of the calorimeter and constructed from two scintillator counters with both a graphite absorber and a tungsten radiator in front of each counter. 

The tracking spectrometer is built from three tracking stations, each with three layers of double-sided silicon microstrip detectors, interleaved with two 1 m-long 0.57~T dipole magnets. The tracker sensors are Semi-Conductor Tracker (SCT) barrel modules from the ATLAS experiment~\cite{Abdesselam:2006wt}, which have a hit position resolution of about 20~$\mu$m in the precision coordinate, and about 0.6~mm in the other coordinate. Each tracker plane contains eight SCT modules, arranged as a 24 $\times$ 24~cm$^2$ square in the transverse plane. The magnets bend charged tracks in the vertical direction, corresponding to the precision coordinate of the tracker. The FASER tracker is described in more detail in Ref.~\cite{FASER:Tracker}. 

The electromagnetic energy of particles is measured by an electromagnetic calorimeter, the most downstream component of the detector. The calorimeter is constructed from four outer ECAL modules from the LHCb experiment~\cite{LHCB:2000ab}. Each module is 12 $\times$ 12~cm$^2$ in the transverse plane, with 66 layers of interleaved 4~mm thick plastic scintillator and 2~mm thick lead plates, corresponding to a total of 25 radiation lengths. A module has 64 wavelength-shifting fibers that penetrate the length of the module and end in a photomultuplier tube (PMT). The readout of the PMTs saturates for large pulses corresponding to energy deposits above 3~TeV. From July to August 2022 the readout was set to saturate at 300~GeV for commissioning purposes, and these data are excluded from the dark photon search. The calorimeter energy resolution has been measured with high energy electrons in a testbeam, provided by the CERN SPS and carried out in July 2021~\cite{testbeampaper}, to be $\mathcal{O}{(1\%)}$ in the high-energy range most relevant for this analysis.

Readout is triggered by signals from the scintillators or calorimeter system, with a typical trigger rate of 1~kHz dominated by high energy muons from IP1. The average detector deadtime was 1.3\%, which is accounted for when calculating the luminosity collected by FASER. The trigger and data acquisition systems are described in more detail in Ref.~\cite{FASERTDAQ:2021}. 

\section{Dataset and Simulation Samples}
\label{sec:DataSim}
This search uses 27.0~fb$^{-1}$ of Run 3 collision data collected by FASER between September and November 2022. The luminosity of the dataset is provided by the ATLAS experiment~\cite{DAPR-2021-01,Avoni:2018iuv, ATL-DAPR-PUB-2023-001} and corrected for the FASER detector dead time.

Monte Carlo (MC) simulation samples are used to evaluate the signal efficiency, in the estimation of background yields, and to calculate the systematic uncertainties. All samples are simulated using \texttt{GEANT4}~\cite{G4} with a perfectly aligned and detailed description of the detector geometry, including passive material. The samples include a realistic level of detector noise, and are reconstructed in the same way as the data.

Signal events are generated using \texttt{FORESEE}~\cite{Kling:2021fwx} with the \texttt{EPOS-LHC}~\cite{Pierog:2013ria} generator to model very forward $\pi^0$ and $\eta$ meson production in the LHC collisions.
The production of dark photons via dark bremsstrahlung is also included, which is modelled using the Fermi-Weizsacker-Williams approximation following Ref.~\cite{Blumlein:2013cua} with the additional requirement on the dark photon's transverse momentum, $p_{\mathrm{T}}(A')<1$~GeV, to ensure the validity of the calculation. Signal samples for the $A'$ and $A'_{B-L}$ models are generated covering the relevant ranges in both coupling and mass. 

A high-statistics high-energy muon sample with $2 \times 10^8$ events entering FASER from IP1 is used for several background and systematic uncertainty studies. The sample uses the expected energy and angle of the muons as estimated by \texttt{FLUKA}~\cite{BATTISTONI201510, Ferrari:2005zk, Bohlen:2014buj} simulations of incoming muons from IP1. The samples include a detailed description of the LHC components and infrastructure between IP1 and FASER. A similar sample of $8 \times 10^5$ large-angle (15-60 mrad) muon events generated slightly upstream of the VetoNu scintillators, and with a radius spanning 15-30~cm covering the edge region of the scintillators, is produced and used to study the background from large-angle muons that miss the veto system.

Neutrino interactions in FASER~\cite{FASER:2023zcr} are simulated by the \texttt{GENIE}~\cite{Andreopoulos:2009rq,Andreopoulos:2015wxa} generator, following the fluence, energy spectrum and flavour composition obtained in Ref.~\cite{Kling:2021gos}. The sample used for the neutrino background study corresponds to 300~ab$^{-1}$ of data, and only includes neutrino interactions upstream of the Veto scintillators and in the active detector area.

\section{Event Reconstruction}
\label{sec:reco}
Event reconstruction is performed using FASER's Calypso~\cite{calypso} offline software system, based on the open-source Athena framework~\cite{ATL-PHYS-PUB-2009-011,athena} from the ATLAS experiment. Charged particle track reconstruction is performed using the combinatorial Kalman filter from the ACTS library~\cite{ACTS}.
When reconstructing multiple tracks, it is required that they do not share more than 6 clusters of hits in contiguous silicon strips on each side of an SCT module; if the number of shared hits exceeds this threshold, then the track with the higher $\chi^{2}$ is discarded.

A track-based alignment of the tracking detector is performed using an iterative local $\chi^2$ alignment method, and shows an improved agreement in the hit residual and track $\chi^2$ distributions when comparing to the perfectly aligned MC. The alignment only considers the most sensitive distortions, translations in the precision tracker coordinate (vertical) and rotations around the longitudinal axis, at both the individual module and tracking layer level.

Extraction of the PMT charge from the scintillator and calorimeter modules is done by summing the digitised waveform values after pedestal subtraction.

The calorimeter charge-to-energy scale calibration is determined using high energy electron and muon beams from the testbeam data described in \cref{sec:FASER}. To take into account differences between the detector configurations in the testbeam and in collision data, the most probable calorimeter charge deposited by muons as minimum ionising particles (MIPs) is used as an in-situ normalisation of the energy scale. Special calibration runs are performed at high calorimeter gain to measure the MIP signal. After individually normalising each calorimeter module signal to the MIP scale, the testbeam data are used to estimate the initial electromagnetic energy of the particle entering the calorimeter.

\section{Event Selection}
\label{sec:selection}
The typical $A'$ detector signature, shown in \cref{fig:detector_signal}, provides a unique signature to investigate. Since the $A'$ is weakly interacting, no signal is expected in the veto scintillator systems. The $A'$ can then decay in the decay volume to a very collimated, high momentum, $e^+ e^-$ pair, leaving two closely-spaced oppositely-charged particle tracks in the tracker. The $e^+ e^-$ then leave signals in both the Timing and Pre-shower scintillators as well as a large energy deposit in the calorimeter. There are no significant SM processes that can mimic this signature, allowing for a close-to background-free search.

To avoid unconscious bias affecting the analysis, a blinding procedure is applied to events where there is both no signal in any veto scintillator and the calorimeter energy is above 100~GeV. The event selection, background estimation and systematic uncertainties are then finalised before looking in this signal-dominated region of the data.

The signal region event selection requires the following:
\begin{itemize}
    \item event time is consistent with a colliding bunch at IP1;
    \item no signal in any of the five veto scintillators;
    \begin{itemize} 
        \item required to be less than half that expected from a MIP
    \end{itemize} 
    \item signal in the scintillators that are downstream of the decay volume;
    \begin{itemize} 
        \item required to be compatible with or larger than expected for two MIPs
    \end{itemize} 
    \item two fiducial reconstructed tracks of good quality;
    \begin{itemize} 
        \item a good quality track has a track fit $\chi^2$/(number of degrees of freedom) $<$ 25, at least 12 hits on track, and a momentum $>$ 20 GeV
        \item a fiducial track has an extrapolated position of $<$ 9.5~cm radius at all scintillators and tracking stations
    \end{itemize} 
    \item total calorimeter energy greater than 500~GeV;
\end{itemize} 
The efficiency of this selection on a representative signal model in the parameter space where the analysis is most sensitive ($\epsilon=3\times10^{-5}$, $m_{A'}=25.1$\,MeV) was found to be about 50\% for dark photons that decay in the decay volume (where the probability of the dark photon to decay while within the volume is $~\mathcal{O}(10^{-3})$), with the largest inefficiency arising from the two track requirement.

A requirement that the Timing scintillator trigger fired ensures that the trigger efficiency, measured using orthogonal triggers on two-track events, is 100\% for the $A'$ phase space of interest.

The probability to veto a signal event, due to the presence of an uncorrelated beam-background muon in the same or neighbouring bunch crossing, is estimated to be less than 1 per mille.

\section{Backgrounds}
\label{sec:backgrounds}

Several sources of background are considered in the analysis. The dominant background arises from neutrino interactions in the detector. Other processes such as neutral hadrons, or muons that enter the detector volume without firing the veto scintillator systems, either by  missing the scintillators or due  to scintillator inefficiencies, also contribute to the background.
Inefficiencies in the veto scintillators can lead to an instrumental background from unvetoed muons entering the detector volume. Finally, non-collision backgrounds from cosmic-rays or nearby LHC beam interactions are also considered. The contribution of each of these background sources is described and quantified in the following sub-sections.

\subsection{Background Due to Veto Inefficiency}

The inefficiency of each of the five planes of veto scintillators is measured independently with data, by selecting events in which there is a single good fiducial reconstructed track and then measuring the fraction of such events in which the scintillator charge is below that of a MIP signal. Thanks to the thick scintillators and tight fiducial track requirements, the inefficiencies are at the $10^{-5}$ level or smaller. Since the planes are independent, this leads to a combined veto inefficiency of smaller than $10^{-20}$. As $\mathcal{O}{(10^8)}$ incoming muons are observed in the 2022 dataset, the background due to the veto inefficiency is taken to be negligible.

\subsection{Background from Neutral Hadrons}

Neutral hadrons produced in muon interactions in the rock in front of FASER can be a possible source of background if, when passing through the veto systems undetected and interacting or decaying inside the detector decay volume, they produce exactly two reconstructed charged particle tracks and a calorimeter energy deposit above 500~GeV. This background is heavily suppressed by the need for the neutral hadron to traverse the full eight interaction lengths of the FASER$\nu$ detector, and by the need for the parent muon to scatter to miss the veto scintillators.  

To determine the fraction of neutral hadron events that deposit at least 500~GeV of energy in the calorimeter, a three-track control region is used, where the parent muon enters the detector and is reconstructed along with the neutral hadron decay products.  In these three-track events, the ratio of events with low calorimeter energy ($E<100$~GeV) to high energy ($E>500$~GeV) is used to scale the number of events with two reconstructed tracks (in which the parent muon is not present in the detector) at low-energy ($E<100$~GeV) to estimate the expected background number of two-track events with $E>500$~GeV. To allow sufficient event counts in the two-track low-energy control region, the veto requirements are relaxed, requiring no signal in the VetoNu scintillators, but with no requirements on the other Veto scintillator signals.  

Photon conversion events (with the accompanying parent muon) constitute a significant fraction of the three-track sample defined above and must be removed. This is done by requiring that the invariant mass of the two lowest momentum tracks, where the muon is assumed to be the highest momentum track, is greater than 200~MeV, which was found to be optimal when separating $K_S$ events from photon conversions in MC simulation.

After discarding the photon conversion events, the number of data events in the low- and high-energy three-track regions are 404 and 19 respectively. This ratio is used to extrapolate the one event observed in the low-energy two-track region to the two-track high-energy region, resulting in an estimate of 0.047 expected events. This method provides an estimate of the number of neutral hadron events that lead to two reconstructed tracks, contain more than 500~GeV of calorimeter energy, and leave no signal in the VetoNu scintillators.

To obtain the final background estimate, the results are corrected to account for the fact that the signal region selection requires no signal in the downstream Veto station as well. The correction is derived by studying the signal recorded in the Veto station using three-track events. With a clear separation in the Veto scintillator signal size for when only one track (the parent muon) traverses the Veto station versus when the other two tracks also leave a signal in the Veto station, the scintillators can be used to select both types of events and the ratio of the number of events in the two cases is used as the correction.

After correcting for the fraction of events that will decay or interact before the second veto system, a final estimate of $(8.4 \pm 11.9) \times 10^{-4}$ events is found; where the 100\% statistical uncertainty is driven by the single event observed in the low-energy two-track data region, and an additional 100\% systematic uncertainty is applied to account for the assumptions in the method. In performing this estimation, potential neutrino background to the low-energy two-track data region, predicted to be $3.6 \pm 3.8$ events from GENIE simulation, is conservatively neglected.

\subsection{Background from Large-Angle Muons}

Another potential background source arises from large-angle muons that miss the veto system and then enter the FASER decay volume. This background is heavily suppressed by the fact that the tracks extrapolated to the front veto scintillators are required to be within the fiducial volume. The MC sample with large-angle muons generated at the edge of the scintillators, described in \cref{sec:DataSim}, is used to study this background. No two-track events are seen in this sample, even before applying the fiducial requirements on the extrapolated tracks or the calorimeter energy requirement, suggesting that this background is negligible in the final analysis.

This was validated via a data-driven method by using events with a signal in the veto scintillators and calculating the ratio of the number of such events with $>500$\,GeV or $<$ 500~GeV in the calorimeter, which is then used to extrapolate from the number of events with no signal in the veto scintillators and $<$ 500~GeV in the calorimeter to the number of events with no signal in the veto and $>$ 500~GeV in the calorimeter. The results of this validation are consistent with those from the MC estimate, providing confidence that this background is negligible. 

\subsection{Background from Neutrinos}

The large flux of high energy neutrinos, whose interaction cross section rises with energy, at the FASER location constitutes an important background, since the neutrinos do not leave any signal in the veto scintillators, and can interact to produce high energy particles. To suppress this background, the detector was designed to minimise the amount of material in the main detector volume.  

The expected background from neutrino interactions inside the detector is estimated using the 300~ab$^{-1}$ ($\sim10000\times$ larger than the data used in this analysis) neutrino MC sample described in \cref{sec:DataSim}. The MC simulation shows that 0.0015 neutrino events (0.0012 electron (anti)neutrino events and 0.0003 muon (anti)neutrino events) pass the signal region selection when scaled to 27.0~fb$^{-1}$ of data, with these interactions occurring in the Timing scintillator station or the first tracking station. 
\cref{fig:neutrinoEnergy} shows the calorimeter energy distribution for neutrino events that pass the signal region selection when disregarding the requirement on the calorimeter energy. The figure shows that a requirement of $\geq$ 500~GeV gives a good suppression of the neutrino background.
The uncertainty on the incoming neutrino flux~\cite{Kling:2021gos} is taken to be 100\% for electron neutrinos and 25\% for muon neutrinos, and an additional 100\% uncertainty is applied to account for the effect of uncertainties in the modelling of neutrino interactions. The total neutrino background estimate when scaled to 27.0~\fb$^{-1}$ is (1.5 $\pm$ 0.5 (stat.) $\pm$ 1.9 (syst.))$\times10^{-3}$ events.

\begin{figure}[tbp]
\begin{center}
\includegraphics[width=0.7\textwidth]{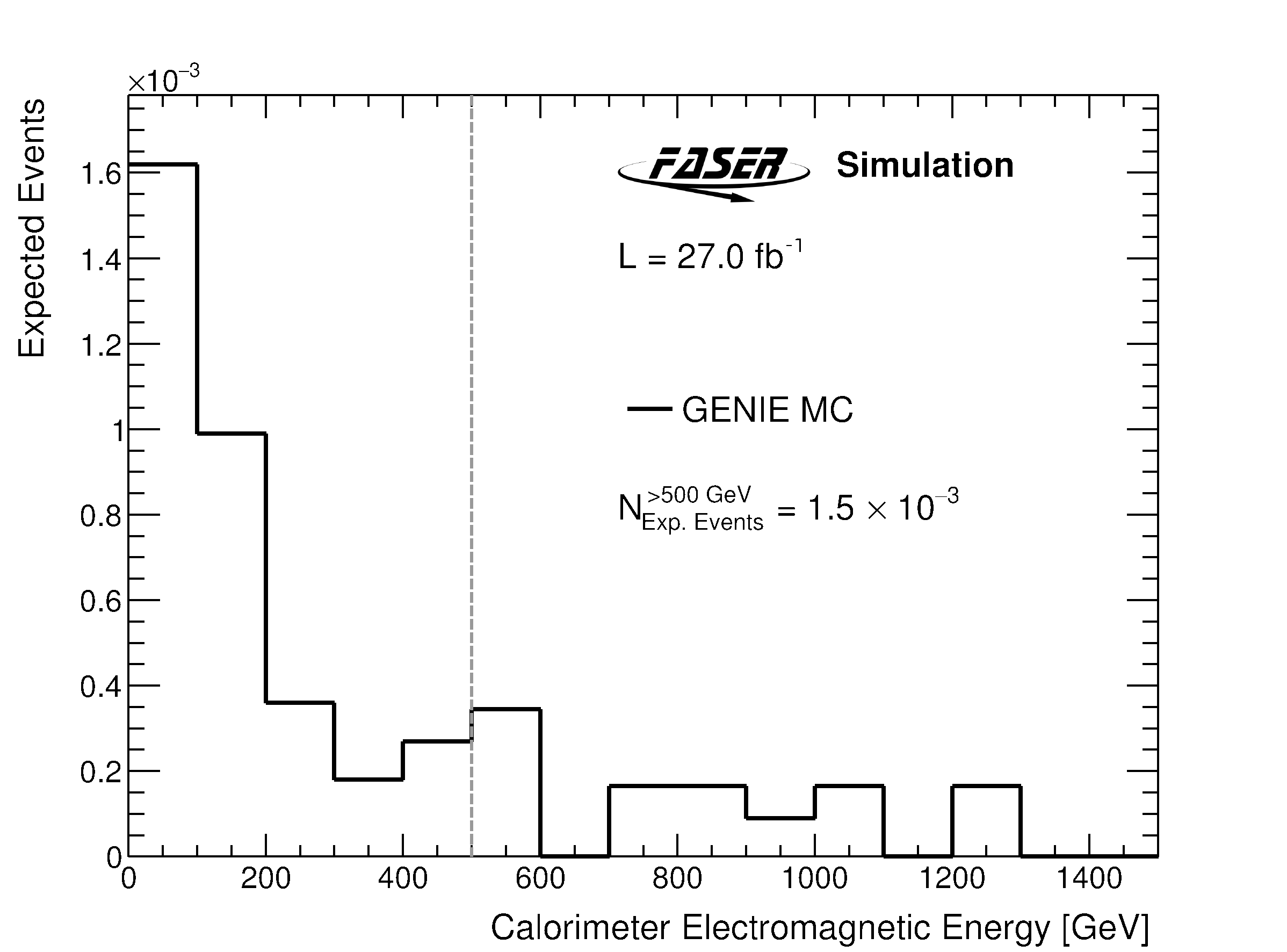}
\caption{The calorimeter energy in simulated neutrino events passing all signal selection requirements, besides that on the calorimeter energy. GENIE is used to simulate the neutrino interactions. The figure is scaled to a luminosity of 27.0~fb$^{-1}$.}   
\label{fig:neutrinoEnergy}
\end{center}
\end{figure}

\subsection{Background from Non-collision events}
The background from cosmic rays and the non-colliding beam background are considered by studying events collected at times when there are no colliding bunches in IP1. Cosmic rays are studied during 330 hours of data-taking with no beam in the machine, which corresponds to a similar running time to the full 2022 physics data-taking period. During this time, no event is observed with a calorimeter energy deposit above 100~GeV, and no events are found when requiring at least one good quality track. 

The beam background from LHC beam-1, the incoming beam to ATLAS in the FASER location, is the most relevant for FASER. Beam-1 interactions with gas or tails of the beam interacting with the beampipe aperture can lead to particles boosted in the direction of FASER, where low-energy activity is observed in correlation with beam-1 bunches passing the back of the detector. This beam background is studied by checking the detector activity in events with the relevant bunch timing, but which do not correspond to colliding bunches at IP1. It is found that beam background events without signal in the veto scintillators do not have a good reconstructed track, and for these events without a track, there are zero events with calorimeter energies above 400~GeV.

The overall contribution from non-collision backgrounds is therefore considered to be negligible. 

\subsection{Summary of the Expected Background}

As background contributions from the veto inefficiency, large-angle muons, and non-collision events are estimated to provide a negligible contribution in the signal region, the total expected background is obtained by combining just the neutrino and neutral hadron estimates, leading to a total background of $(2.3 \pm 2.3) \times 10^{-3}$ events.

\section{Systematic Uncertainties on the Signal Yield}
\label{sec:Systematics}

Systematic uncertainties on the expected signal yields arise from several sources. The uncertainty in the integrated luminosity is provided by the ATLAS collaboration, and is 2.2\%~\cite{ATL-DAPR-PUB-2023-001}, following the methodology discussed in Ref.~\cite{DAPR-2021-01}. The statistical uncertainty from the number of MC simulated signal events is included and ranges from 1 to 3\%. Spin correlations between production and decay are not included in the MC simulated signal, but their effect on this search is negligible~\cite{spincorrelations}. The remainder of the systematic uncertainties, discussed below, arise from the signal generator and from the modelling of the detector response in the MC simulation.

Uncertainties on the number of signal events decaying inside the FASER decay volume are derived by comparing the estimates from using different event generators to model very forward $\pi^0$ and $\eta$ meson production in the LHC collisions. Comparing signal yields from \texttt{QGSJET~II-04}~\cite{Ostapchenko:2010vb} and \texttt{SIBYLL~2.3d}~\cite{Riehn:2019jet} with the central estimate from the \texttt{EPOS-LHC}~\cite{Pierog:2013ria} generator, where these generators have been validated using LHCf's forward photon measurements~\cite{LHCf:2017fnw}, provides an envelope of estimates as a function of the energy of the signal ($E(A')$), that is parameterized and used as the uncertainty:
\begin{equation}
 \frac{\Delta N}{N} =\frac{0.15 + (E(A')/4~\tev)^3}{1 + (E(A')/4~\tev)^3} \ .
 \label{eq:sys_signal_f}
\end{equation}
The parameterization also envelops the uncertainty on the signal predictions due to changing the $p_\mathrm{T}$-cutoff in modelling of the dark bremsstrahlung as described in \cref{sec:DataSim}.
\cref{fig:signalSysts} presents the $A'$ energy distribution as estimated by the different generators for a representative signal model. The parameterisation is checked for numerous signal models spanning the relevant phase space for both the $A'$ and $A'_{B-L}$ gauge bosons, and is found to be in good agreement with the envelope of the generators.

\begin{figure}[tbp]
\begin{center}
\includegraphics[width=0.7\textwidth]{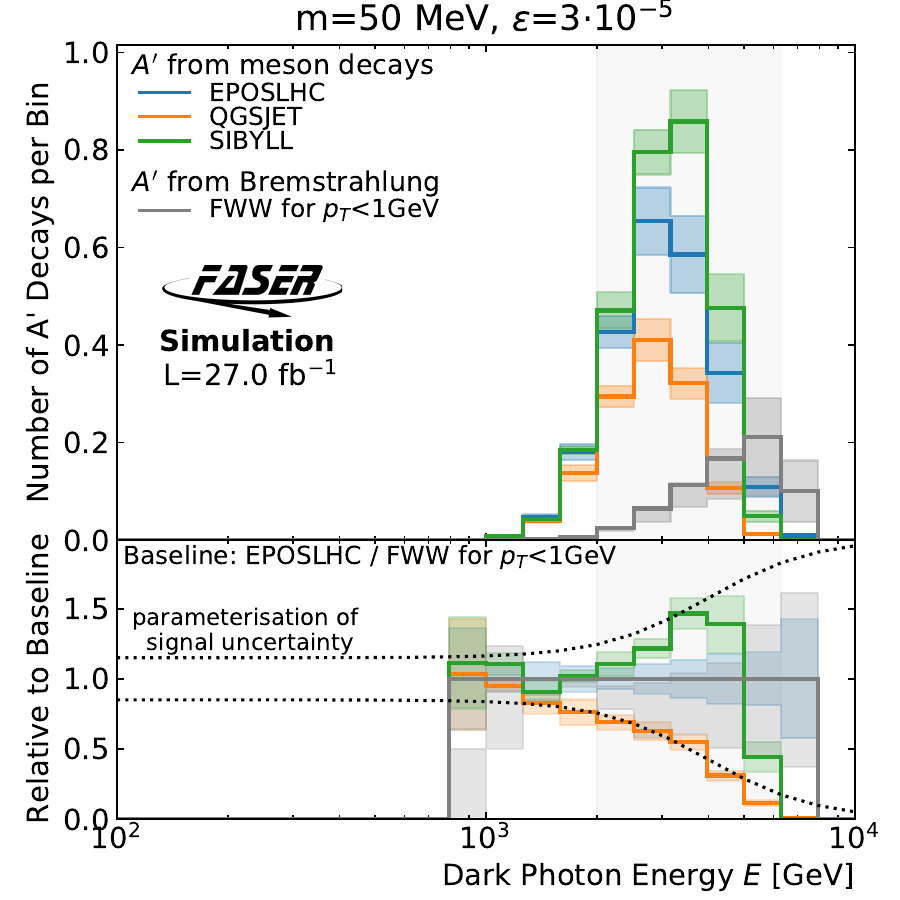}
\caption{The energy spectrum of dark photons in FASER produced with meson production modeled by different generators (\texttt{EPOS-LHC}, \texttt{QGSJET~II-04} and \texttt{SIBYLL~2.3d}). Also shown is production from bremsstrahlung with a factor of two variation in the $p_{\mathrm{T}}$ cut off. The bottom panel shows the ratio between the different estimates, and the parameterisation of the uncertainty as a function of energy. A representative signal model \\(with m$_{A'}$=50~MeV and $\epsilon$=3 $\times$ 10$^{-5}$) is shown. }   
\label{fig:signalSysts}
\end{center}
\end{figure}

The remaining uncertainties arise from the modelling of the detector response in the MC simulation, which is used to calculate the signal yield.

The scintillator efficiencies are measured to be 100\% in both data and MC simulation for the $A'$ signal, based on the scintillator PMT charge observed in events with two reconstructed tracks, thus no corresponding uncertainty is assigned.

The calorimeter energy scale calibration, as described in \cref{sec:reco}, is applied to both the data and MC simulation identically. The stability of the calorimeter system across the data taking period is tested with regular calibrations using an LED pulse injected into the calorimeter modules~\cite{FASER:2022hcn}. A conservative analysis taking into account all components of the energy calibration leads to a 6\% uncertainty on the difference in the calibration of the energy scale between data and MC simulation. This uncertainty is checked in data by using the $E/p$ distribution in three-track events, which are dominated by photon conversions initiated by high-energy muons. Only the two lowest momentum tracks are considered when calculating the $E/p$ since the highest momentum track is assumed to be the muon. The reconstructed $E/p$ peak position in data and MC simulation is consistent and well within the 6\% uncertainty across the momentum range probed, as shown in \cref{fig:EoP}.

\begin{figure}[tbp]
\begin{center}
\includegraphics[width=0.7\textwidth]{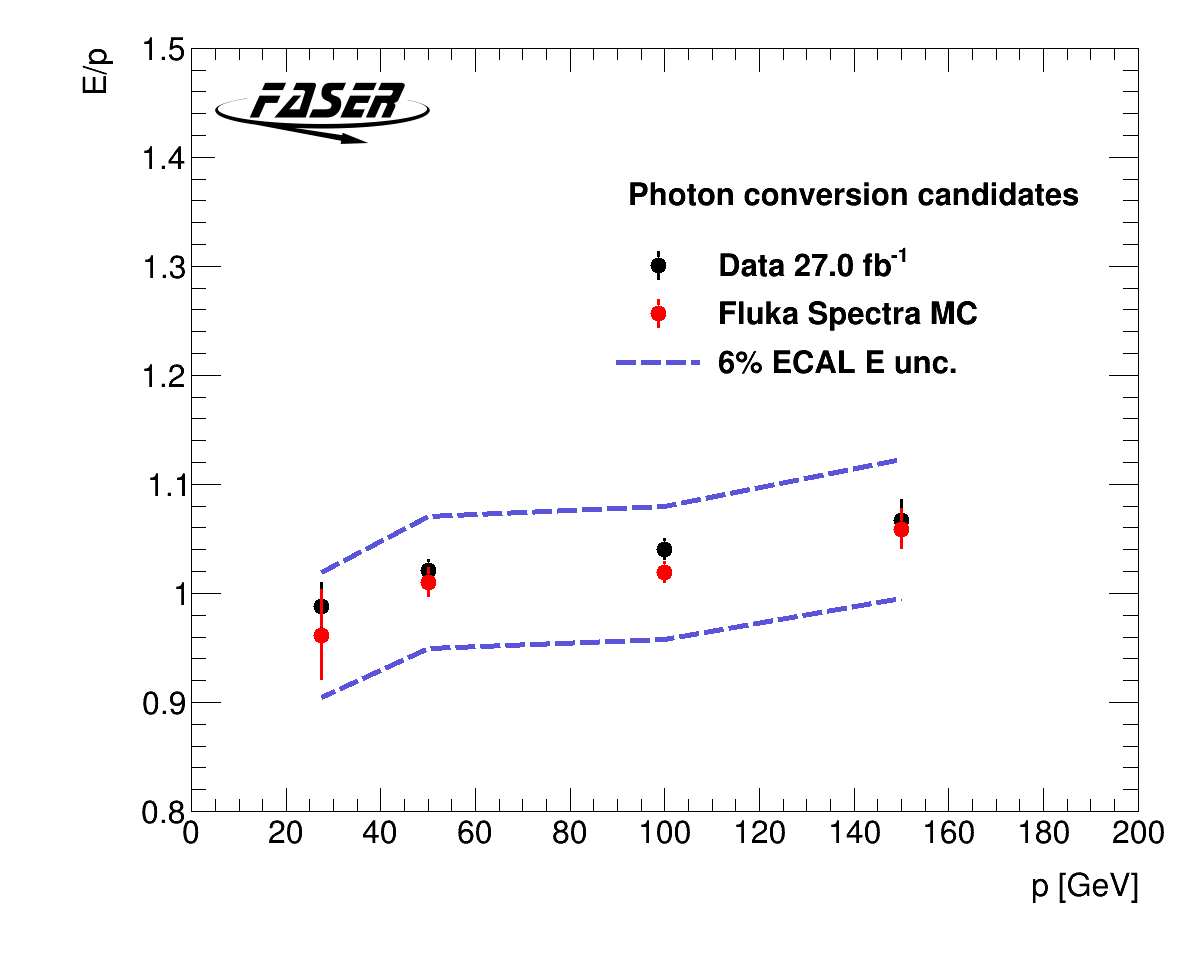}
\caption{The Gaussian-fitted peak position of the $E/p$ in data and MC simulation as a function of the momentum of photon conversion candidates.}   
\label{fig:EoP}
\end{center}
\end{figure}

The uncertainty due to the tracking efficiency of single tracks is assessed by comparing the relative efficiency for finding tracks in events with a single track segment in each of the three tracking stations between data and MC simulation. This yields a 1.5\% uncertainty per track.

The track finding procedure is more complex when there are two closely-spaced tracks, as in the signal, in particular when the tracks share hits. The uncertainty due to this is assessed by overlaying the raw tracker data from two different events, each of which has a single reconstructed track. The track reconstruction is then re-run on the combined event, built from the two overlaid events, so that the tracking efficiency can be calculated. This is performed using both data and simulation, shown in \cref{fig:overlayEfficiency}, where the ratio of the efficiency between the two, as a function of the distance between the two tracks at their first measurements, is used to assess the uncertainty. The efficiency in data is up to 7\% less than in MC simulation, at track separations comparable to that expected in the $A'$ signals, hence a 7\% correction to the two-track tracking efficiency is applied, with a corresponding systematic uncertainty, assumed to be the difference between the nominal and corrected efficiency, applied.

\begin{figure}[btp]
    \centering
    \includegraphics[width=0.7\textwidth]{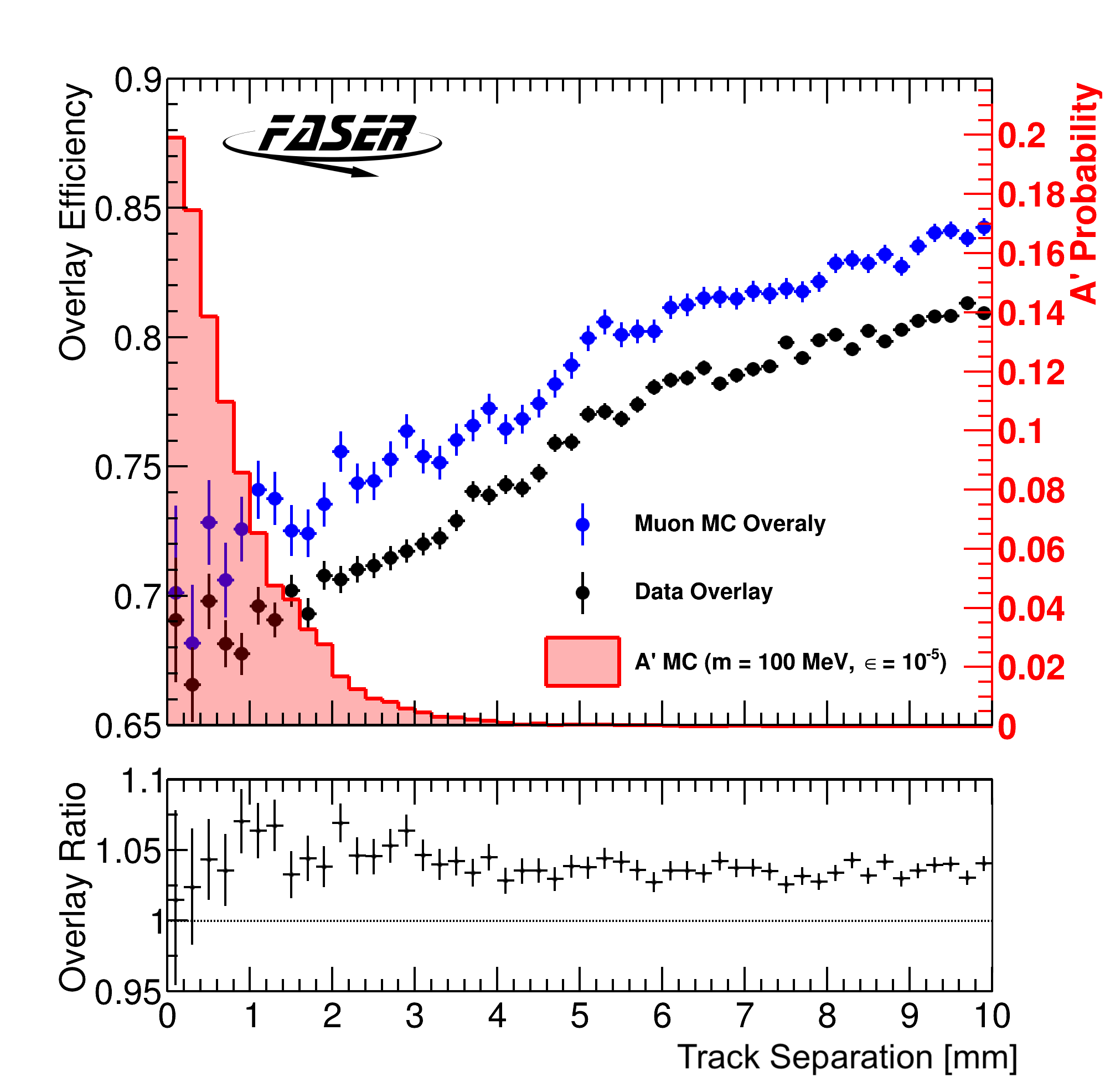}
    \caption{(Top) The two-track reconstruction efficiency versus track separation for overlaid tracks in both data and MC events are shown. The distribution of the separation in $e^+e^-$ tracks of an $A'$ sample is also shown in red with the axis on the right-hand side. (Bottom) The ratio of the overlay tracking efficiencies between MC and data is depicted.}
    \label{fig:overlayEfficiency}
\end{figure}

The track momentum scale and resolution uncertainty is derived by comparing the mass peak of photon conversion events in data and MC simulation. Upon comparison, both a shift or Gaussian smear of the MC track momentum by 5\% were shown to more than account for the difference in the photon conversion mass peak position between data and MC, leading to a conservative uncertainty of 5\% on both the track momentum scale and momentum resolution.

Table~\ref{tab:systs} summarises the various sources of uncertainty on the signal, showing the size of the individual uncertainties, and the range of the effect on the overall uncertainty on the signal yield.

\begin{table}[tbh]
    \centering
    \begin{tabular}{|c|c|c|}    
    \hline
    Source & Value & Effect on signal yield \\
    \hline
    
    Signal Generator & $\frac{0.15 + (E_{A'} / 4 \text{TeV})^3}{1 + (E_{A'} / 4 \text{TeV})^3}$ & 15-65\% (15-45\%) \\
    Luminosity & 2.2\% & 2.2\% \\
    MC Statistics & $\sqrt{\sum{W^2}}$ & 1-3\% (1-2\%) \\

    Track Momentum Scale & 5\% & $<$ 0.5\% \\
    Track Momentum Resolution & 5\% & $<$ 0.5\% \\
    Single Track Efficiency & 3\% & 3\% \\
    Two-track Efficiency & 7\% & 7\% \\ 

    Calorimeter Energy Scale & 6\% & 0-8\%  ($<$ 1\%)\\
    
    \hline    

\end{tabular}
\caption{Summary of the systematic uncertainties on the signal yield. For each of the sources of uncertainty, the source and size of the uncertainty is presented. The effect on the signal yield across the full signal parameter space probed is also shown. The numbers in parenthesis indicate the effect on the signals within the parameter space for which this analysis is sensitive.}
\label{tab:systs}
\end{table}

\section{Results}
\label{sec:results}

After applying the signal selection described in \cref{sec:selection}, zero events are observed in the data, which is compatible with the expected background of (2.3 $\pm$ 2.3) $\times 10^{-3}$ events. \cref{fig:selectionPlots} shows the calorimeter energy distribution for data and three representative signal models at different stages of the signal region selection on the veto scintillator and track information. There are events that have no veto signal and at least one track, but the calorimeter energies are well below the 500~GeV threshold; and there are no events upon further requiring two fiducial tracks.

\begin{figure}[tbp]

\subfigure[]{
\includegraphics[width=0.49\textwidth]{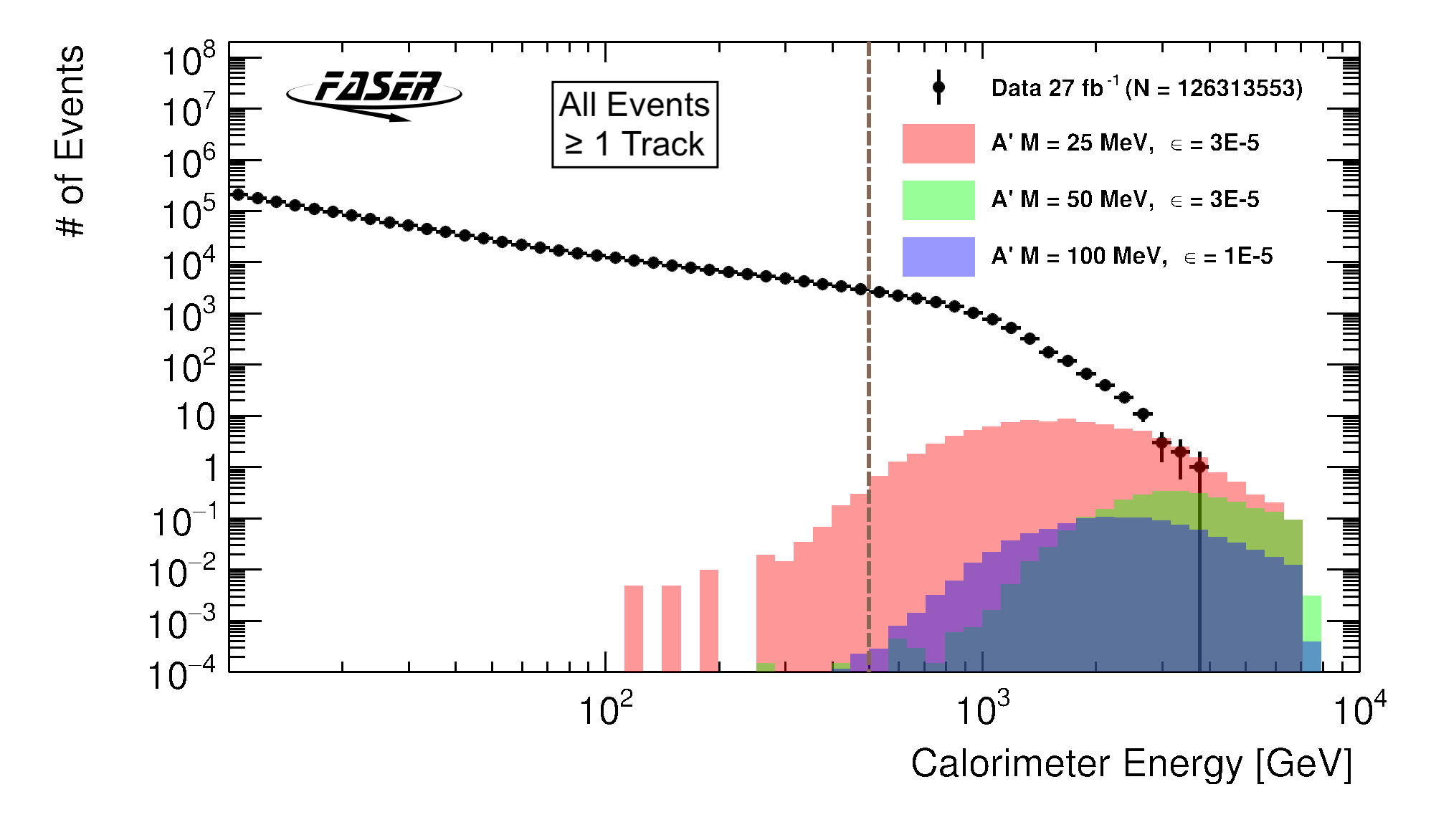}
\label{fig:selectionPlots-a}
}\subfigure[]{
\includegraphics[width=0.49\textwidth]{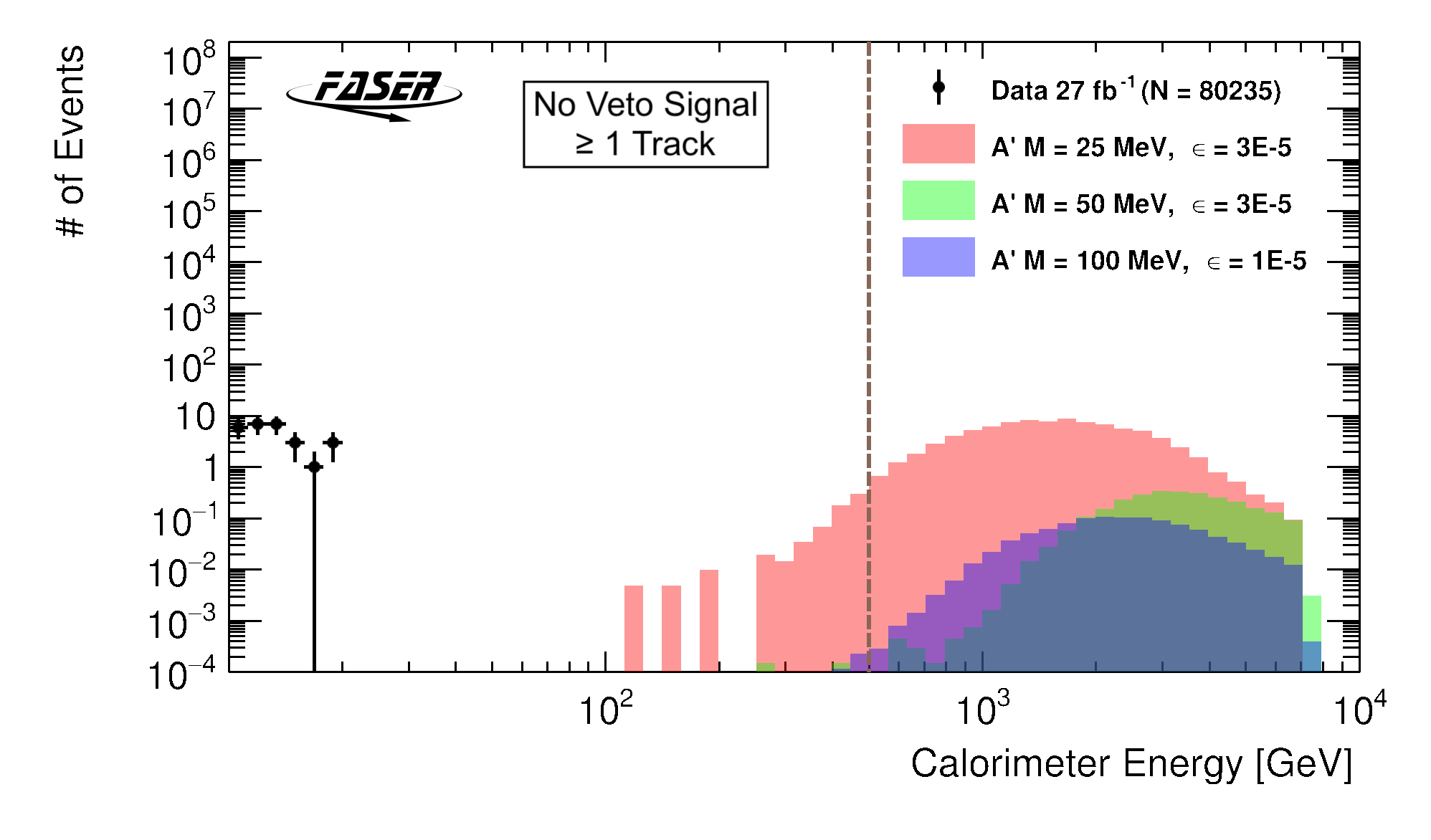}
\label{fig:selectionPlots-b}
}
\subfigure[]{
\includegraphics[width=0.49\textwidth]{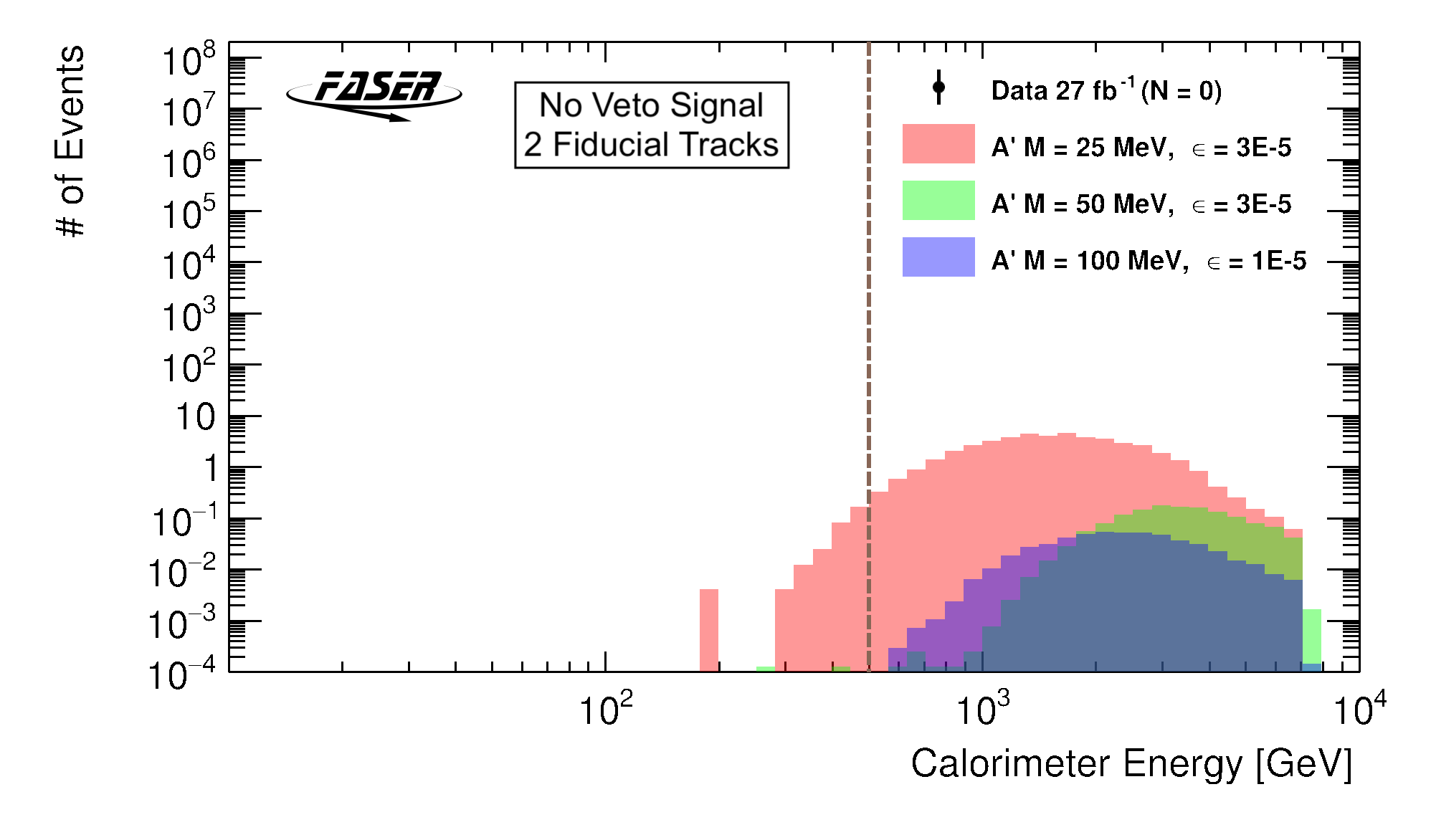}
\label{fig:selectionPlots-c}
}

\caption{The calorimeter energy distribution for data and three representative MC simulated signal models are shown for (a) all events with at least one good track, (b) events that have no signal in the veto stations and at least one good track, and (c) events that have no signal in the veto stations and exactly two good fiducial tracks. The distributions and expected events from the MC samples are scaled to 27.0~fb$^{-1}$.}   
\label{fig:selectionPlots}
\end{figure}

As no significant excess of events over the background is observed, the results are used to set exclusion limits in the signal scenarios considered. The exclusion limits are made using a profile likelihood approach implemented via the \texttt{HistFitter} framework~\cite{Baak:2014wma}, and are set at 90\% confidence level to allow for direct comparison with constraints from other experiments. Hypothesis tests are performed using profile likelihood test statistics~\cite{Cowan:2010js} and the CLs method~\cite{CLsMethod:2002Read} to test the exclusion of new physics scenarios.
For dark photons, the analysis excludes signal models in the range $\epsilon \sim 4 \times 10^{-6} - 2 \times 10^{-4}$ and $m_{A'} \sim 10~\mev - 80~\mev$, and provides the world-leading exclusion for scenarios in the range $\epsilon \sim 2 \times 10^{-5} - 1 \times 10^{-4}$ and $m_{A'} \sim 17~\mev - 70~\mev$. \cref{fig:exclusion_limit-A'} shows the $A'$ exclusion limit in the signal parameter space, where the grey regions are already excluded by experimental data from BaBar~\cite{BaBar:2014zli}, E141~\cite{Riordan:1987aw}, NA48~\cite{NA482:2015wmo}, NA64~\cite{NA64:2019auh}, Orsay~\cite{Davier:1989wz, Andreas:2012mt}, and NuCal~\cite{Blumlein:1991xh, Blumlein:2013cua}, which are adapted from DarkCast~\cite{Ilten:2018crw}. 

A key reason for investigating dark photons is their potential as intermediaries between the SM and a dark sector. In particular, they allow for obtaining the correct value of the dark matter relic density, $\Omega_\chi^{\textrm{total}} h^2\simeq 0.12$~\cite{Planck:2018vyg}, via the thermal freeze-out mechanism. In \cref{fig:exclusion_limit-A'}, an example thermal relic contour is included, obtained for the scenario where the dark photons couple to a light complex scalar dark matter field $\chi$~\cite{Kling:2021fwx}. In particular, this line assumes that the mass ratio between the dark matter candidate and the dark photon is always equal to $m_\chi / m_{A'} = 0.6$  and that the dark photon coupling constant to dark matter has a fixed value of $\alpha_D = 0.1$. This mass ratio guarantees that the dark photon decays visibly into the SM species and that the dark matter primarily annihilates via $\chi\chi \to A' \to ff$. Variations of both the coupling and mass ratio in the dark sector are possible and will lead to a shift of the relic target line. Notably, in the context of this particular dark matter model, the region below the target line would have an over-abundance of dark matter and would be excluded cosmologically: FASER therefore probes a significant fraction of the cosmologically-allowed region of parameter space.

\begin{figure}[tbp]
\subfigure[]{
    \includegraphics[width=0.49\textwidth]{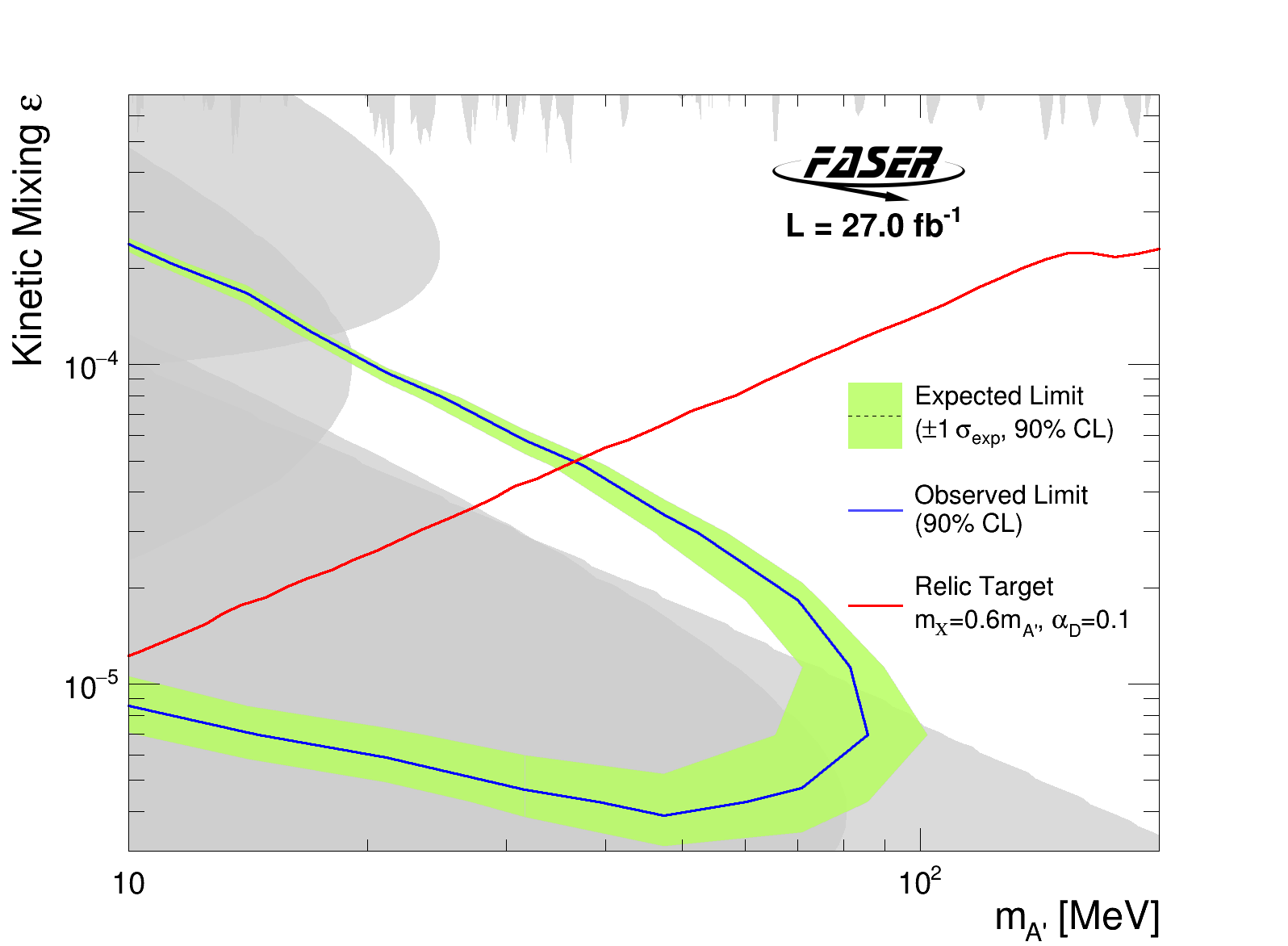}
    \label{fig:exclusion_limit-A'}
}\subfigure[]{
    \includegraphics[width=0.49\textwidth]{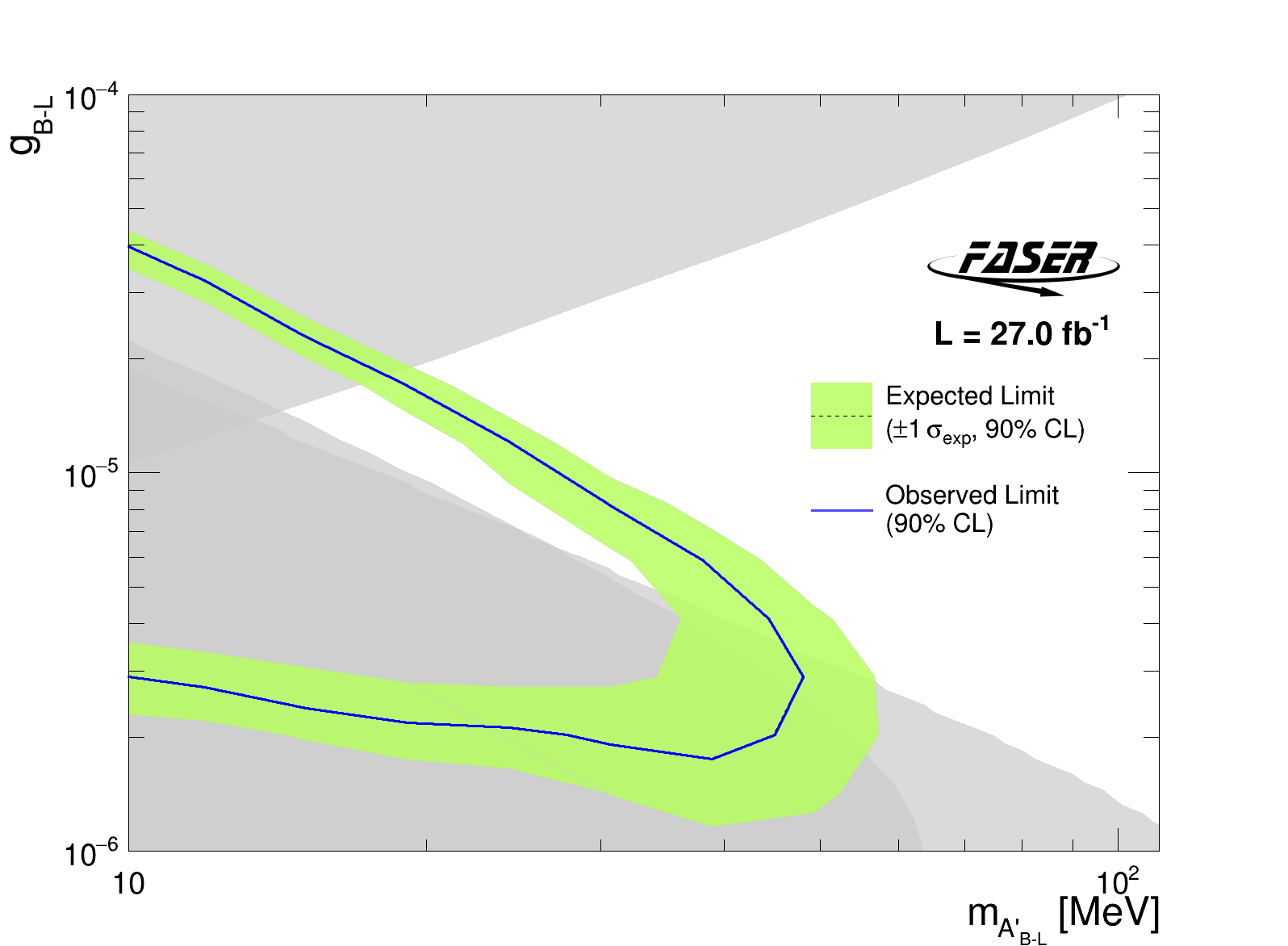}
    \label{fig:exclusion_limit-B-L}
}
\caption{90\% confidence level exclusion contours in (a) the dark photon and (b) the $B-L$ gauge boson parameter space are shown. Regions excluded by previous experiments are shown in grey. The red line shows the region of parameter space that yields the correct dark matter relic density, with the assumptions discussed in the text.} 
\label{fig:exclusion_limit}
\end{figure}
The exclusion contours for the $B-L$ gauge boson are shown in \cref{fig:exclusion_limit-B-L}, where FASER provides the first exclusion for models in the range $g_{B-L} \sim 5\times 10^{-6} - 2\times 10^{-5}$ and $m_{A'_{B-L}} \sim 15~\mev - 40~\mev$, with a total region between $g_{B-L} \sim 3\times 10^{-6} - 4\times 10^{-5}$ and $m_{A'_{B-L}} \sim 10~\mev - 50~\mev$ excluded. In grey are the regions already excluded by experimental data from Orsay~\cite{Davier:1989wz, Andreas:2012mt} and NuCal~\cite{Blumlein:1991xh, Blumlein:2013cua} as adapted from DarkCast~\cite{Ilten:2018crw}, as well as from a dedicated search for invisible final states by NA64~\cite{NA64:2022yly}. In this model, the region probed by FASER is also cosmologically relevant.  Assuming a dark matter particle $\chi$ with a mass in the range of $0.5\times m_{A'_{B-L}}<m_{\chi}<m_{A'_{B-L}}$ and a very large $B-L$ charge, the region of parameter space favored by thermal freeze-out includes regions of parameter space that are now excluded by the new FASER constraint~\cite{Berlin:2018bsc, Mohapatra:2019ysk}. Alternatively, since the $B-L$ model necessarily includes 3 sterile neutrinos, it is natural to consider the possibility that these sterile neutrinos are the dark matter. These sterile neutrinos may be produced through the freeze-in mechanism, and the resulting relic density may be significant in the regions of parameter space probed by FASER~\cite{Kaneta:2016vkq, Mohapatra:2019ysk, Eijima:2022dec}.

\section{Conclusions}
\label{sec:conclusions}
The first search for dark photons by the FASER experiment has been presented, providing a proof of principle that very low background searches for long-lived particles in the very forward region are possible at the LHC. The search applies an event selection requiring no signal in the veto scintillator systems, two good quality reconstructed charged particle tracks and more than 500~GeV of energy deposited in the calorimeter. No events are observed passing the selection, with an expected background of (2.3 $\pm$ 2.3) $\times 10^{-3}$ events. At the 90\% confidence level, FASER excludes the region of $\epsilon \sim 4 \times 10^{-6} - 2 \times 10^{-4}$ and $m_{A'} \sim 10~\mev - 80~\mev$ in the dark photon parameter space, as well as the region of $g_{B-L} \sim 3\times 10^{-6} - 4\times 10^{-5}$ and $m_{A'_{B-L}} \sim 10~\mev - 50~\mev$ in the $B-L$ gauge boson parameter space. In both the dark photon and $B-L$ gauge boson models, these results are one of the first probes of these regions of parameter space since the 1990’s, and they exclude previously-viable models motivated by dark matter.

\section{Acknowledgments}
\label{sec:Acknowledgments}
We thank CERN for the very successful operation of the LHC during 2022. We thank the technical and administrative staff members at all FASER institutions for their contributions to the success of the FASER project. We thank the ATLAS Collaboration for providing us with accurate luminosity estimates for the used Run 3 LHC collision data. FASER gratefully acknowledges the donation of spare ATLAS SCT modules and spare LHCb calorimeter modules, without which the experiment would not have been possible. We also acknowledge the ATLAS collaboration software, Athena, on which FASER’s offline software system is based~\cite{ATL-PHYS-PUB-2009-011} and the ACTS tracking software framework~\cite{ACTS}. Finally we thank the CERN STI group for providing detailed FLUKA simulations of the muon fluence along the LOS, which have been used in this analysis. This work was supported in part by Heising-Simons Foundation Grant Nos. 2018-1135, 2019-1179, and 2020-1840, Simons Foundation Grant No. 623683, U.S. National Science Foundation Grant Nos. PHY-2111427, PHY-2110929, and PHY-2110648, JSPS KAKENHI Grants Nos. JP19H01909, JP20K23373, JP20H01919, JP20K04004, and JP21H00082, BMBF Grant No. 05H20PDRC1, DFG EXC 2121 Quantum Universe Grant No. 390833306, ERC Consolidator Grant No. 101002690, Royal Society Grant No. URF\textbackslash R1\textbackslash 201519, UK Science and Technology Funding Councils Grant No. ST/ T505870/1, the National Natural Science Foundation of China, Tsinghua University Initiative Scientific Research Program, and the Swiss National Science Foundation.

\bibliographystyle{utphys}
\bibliography{references}


\end{document}